\documentclass{aa}
\usepackage{times}
\usepackage[sectionbib,authoryear]{natbib} 
\usepackage{graphicx} 
\newcommand{\gsim}{\,\raisebox{0.2em}{$>$}\!\!\!\!\!
\raisebox{-0.25em}{$\sim$}\,}

\newcommand{\gr}{$\gamma$-ray \,}
\newcommand{\grs}{$\gamma$-rays \,}
\newcommand{\rxj}{RX~J1713.7-3946 \,}

\begin{document}

\title {Nonthermal and thermal
    emission from the supernova remnant RX~J1713.7-3946}

\titlerunning{Nonthermal vs. thermal emission from SNR RX~J1713.7-3946}

   \subtitle{}

   \author{ E.G. Berezhko
          \inst{1}
          \and         
          H.J.V\"olk 
          \inst{2}
          }

   \offprints{H.J.V\"olk}

   \institute{Yu.G. Shafer Institute of Cosmophysical Research and Aeronomy,
                     31 Lenin Ave., 677980 Yakutsk, Russia\\
              \email{berezhko@ikfia.ysn.ru}
         \and
           Max Planck Institut f\"ur Kernphysik,
                Postfach 103980, D-69029 Heidelberg, Germany\\
              \email{Heinrich.Voelk@mpi-hd.mpg.de}             
             }

   \date{Received month day, year; accepted month day, year}

   \abstract{} 
   {A nonlinear kinetic theory of cosmic ray (CR) acceleration in supernova
     remnants (SNRs) is employed to investigate the properties of SNR
     RX~J1713.7-3946.}
   {Observations of the nonthermal radio and X-ray emission spectra as well as
     the H.E.S.S. measurements of the very high energy \gr emission are used to
     constrain the astronomical and CR acceleration parameters of the
     system. It is argued that \rxj is a core collapse supernova (SN) of type
     II/Ib with a massive progenitor, has an age of $\approx 1600$~yr and is at
     a distance of $\approx 1$~kpc. It is in addition assumed that the CR
     injection/acceleration takes place uniformly across the shock surface for
     this kind of core collapse SNR.}  
   {The theory gives a consistent description for all the existing
     observational data, including the nondetection of thermal X-rays and the
     spatial correlation of the X-ray and \gr emission in the remnant.
     Specifically it is shown that an efficient production of nuclear CRs,
     leading to strong shock modification and a large downstream magnetic field
     strength $B_\mathrm{d}\approx 140$~$\mu$G can reproduce in detail the observed
     synchrotron emission from radio to X-ray frequencies together with the \gr
     spectral characteristics as observed by the H.E.S.S. telescopes.}
   {The calculations are consistent with \rxj being an efficient source of
     nuclear cosmic rays.}



   \maketitle
%

\section{Introduction}

RX~J1713-3946 is a shell-type supernova remnant (SNR) located in the Galactic
plane that was discovered in X-rays with {\it ROSAT} \citep{pfe96}.
Subsequent studies of this SNR with the {\it ASCA}
satellite by \citet{koyama97} and  \citet{sla99}, and later
with {\it XMM} by \citet{cassam04}, have not positively identified any
  thermal X-ray emission and therefore tentatively concluded
that the observable X-ray emission is entirely non-thermal.

The radio emission on the other hand is quite weak. Only part of the remnant
shell could be detected in radio synchrotron emission up to now, with a poorly
known spectral form \citep{laz04}. For the spatially integrated radio flux
\citet{aha06} estimated about twice the value found by \citet{laz04}. Most
recently \citet{acero09} derived a total flux value of $22 \mathrm{Jy} <
\mathrm{S} < 26 \mathrm{Jy}$ at 1.4~GHz, still about two times higher than
estimated by \citet{aha06}. As far as the estimate of the magnetic field
strength is concerned, this higher estimate of the radio synchrotron flux
increases the magnetic field estimate (see below).

RX~J1713-3946 was also detected in very high energy (VHE:$ >100$~GeV) \grs with
the {\it CANGAROO} \citep{mur00,enomoto02} and {\it H.E.S.S.}  telescopes
\citep{aha04,aha06,aha07}. Especially the latter, very detailed observations
show a clear shell structure at TeV energies which correlates well with the
{\it ASCA} contours.

In a first theoretical paper on this source \citep{bv06} we have
investigated the acceleration of CR electrons and protons
in detail, using nonlinear kinetic theory \citep{byk96,bv00}. Observations of
the nonthermal radio and X-ray emission spectra as well as the
H.E.S.S. measurements of the very high energy \gr emission were used to
constrain the astronomical and the particle acceleration parameters of the 
system. Under the assumption that \rxj is the remnant of a
core collapse supernova (SN) of type II/Ib with a massive progenitor, the
theory gives indeed a consistent description for the existing data on the
nonthermal emission. Specifically the VHE data from H.E.S.S. were shown to be
best explained as hadronic \gr emission, where the magnetic field amplification
strongly depresses the inverse Compton and Bremsstrahlung fluxes. Subsequently
\citep{bv08}, we have analyzed the spatial correlation between the synchrotron
and the VHE emission that has been used before as an argument for a leptonic
origin of the VHE emission \citep{Katz08,Plaga}. It was
argued that correlated density and magnetic field strength variations lead
rather naturally to a spatial correlation of the hadronic VHE emission with the
synchrotron emission. Similar arguments were brought forward by
\citet{tua08}. Also the recent broad-band X-ray synchrotron measurements
\citep{uat07,ttu08} with the Suzaku instrument were compared to the theoretical
synchrotron spectra and found to be quite consistent with a hadronic model. A
purely leptonic model on the other hand \citep[e.g.][]{Porter06}, where
magnetic field amplification is not expected, was shown to be
inconsistent with these observations. (\citet{acero09} have also considered
  a leptonic origin of the \gr emission as a viable possibility. This has to be
  compared with the results of the present paper.)
  
These theoretical considerations assumed that only over part
of the shock surface injection of suprathermal nuclear ions occurs effectively,
because over other parts the magnetic field is essentially parallel to the
shock surface. In fairly regular SNRs like those of Type Ia explosions into a
presumably uniform circumstellar environment, this is an important effect, as
for instance the case of SN~1006 shows. For a blast wave propagating into the
highly turbulent shell of a stellar wind bubble from a massive progenitor star
on the other hand, the magnetic flux tubes in the shell are presumably so
slender and irregular that energetic particles can cross effusively into
regions where suprathermal injection is depressed, yet acceleration is
possible. As a result shock modification by the pressure gradient of the
accelerating particles becomes possible everywhere on the shock surface. This
reasoning follows the discussion in \citet{bpv09} with the conclusion that SNRs
propagating into wind bubbles should have their forward shock modified
everywhere. Technically this implies that the SNR becomes spherically symmetric
and the correction factor $f_\mathrm{re}<1$ \citep{vbk03} on the spherically
symmetric solution reaches unity. The ubiquitous shock modification has also
implications for the thermal emission, because the thermal gas is then heated
only in the subshock and the average gas density required for a given hadronic
\gr emission is reduced.    

Contrary to the previous studies \citep{bv06,bv08} of \rxj we therefore adopt
here the approximation of spherically symmetric CR injection/acceleration with
$f_{\mathrm{re}} \approx 1$~ \citep[see also][]{zirak09}. It is demonstrated
that also in this approximation a consistent solution can also be found. In
addition a rough estimate of the thermal X-ray emission from this object shall
be given. Even though the error in this estimate might be quite large, the
nominal result is consistent with the present non-detection of thermal X-rays
\citep[e.g.][]{acero09}.  Finally, the observed correlation of the X-ray and
the VHE \gr emission is discussed in a generalized form.

\section{Results}

As in the previous studies \citep{bv06,bv08} a source distance
of $d=1$~kpc is adopted here. For the present angular source size of
about 60 arcmin this implies a SNR radius $R_\mathrm{s}\approx 10$~pc.

For the SNR age the value $t_\mathrm{sn}=1612$~yr is used,
consistent with the hypothesis of \citet{wang97} that RX~J1713-3946 is the
remnant of the AD393 ``guest star''.

It had been argued \citep{bv06} that the existing data are only consistent with
\rxj being a core collapse SN of type II/Ib. Those progenitor stars of core
collapse SNe that emit intense winds are massive main-sequence stars with
initial masses $M_\mathrm{i} \gsim 15M_{\odot}$. During their evolution in a
surrounding uniform ISM of gas number density $\varrho_0=m_pN_\mathrm{ISM}$, the
wind termination shock creates a hot bubble of shocked wind material,
surrounded by a turbulent shell of shocked ISM behind an outer forward shock
that communicates the internal overpressure to the environment.

The gas number density distribution $N_\mathrm{g}(r)=\varrho(r)/m_\mathrm{p}$ is
  assumed to have the form
\begin{equation}
N_\mathrm{g}=0.008+0.24[r/(10~\mbox{pc})]^{12}~~\mbox{cm}^{-3}
\end{equation}
This density profile corresponds to a gas number density
$N_\mathrm{g}=0.25$~cm$^{-3}$ at the current SN shock position, which is a
factor $\sim 5$~lower than in \citet{bv06}.

Under these conditions a SN explosion energy $E_\mathrm{sn}=1.3\times
10^{51}$~erg and an ejecta mass $M_\mathrm{ej}=3M_{\odot}$ lead to a good fit
for the observed SNR properties. To determine the explosion energy directly
from the data the value of the shock speed $V_\mathrm{s}$ would have been
needed. However, this quantity is not measured. The ejecta energy
  $E_\mathrm{ej}$ is still about $0.25 E_\mathrm{sn}$. Together with the
  kinetic energy of the shocked gas 
$E_\mathrm{gk}\approx 0.2 E_\mathrm{sn}$ this makes up about half of the total
  energy $E_\mathrm{sn}$ (see Fig.1.).

\begin{figure}
\centering
\includegraphics[width=7.5 cm]{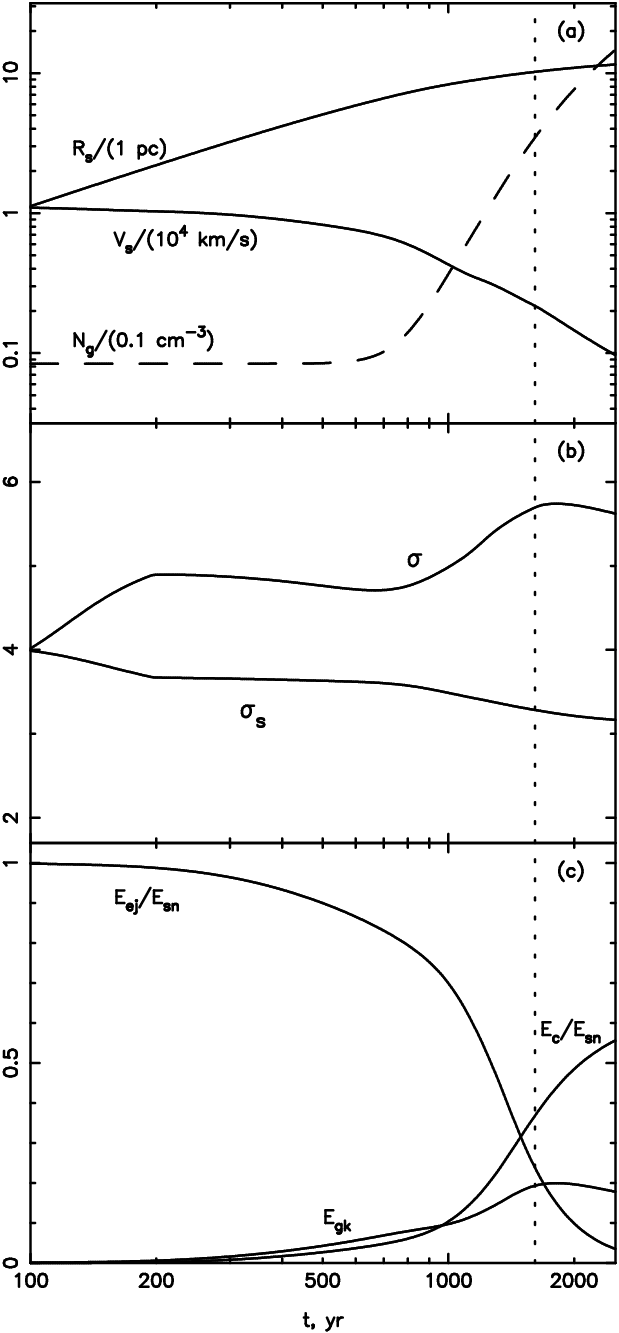}
\caption{(a) Shock radius $R_\mathrm{s}$, shock speed $V_\mathrm{s}$, and
  preshock gas density $N_\mathrm{g}$; (b) total shock ($\sigma$) and subshock
  ($\sigma_\mathrm{s}$) compression ratios; (c) ejecta ($E_\mathrm{ej}$)
gas kinetic ($E_\mathrm{gk}$) and CR
  ($E_\mathrm{c}$) energies as a function of time, normalised to the total
  mechanical SNR energy $E_\mathrm{sn}$. The vertical dotted line marks the
  current evolutionary epoch.}
\label{f1}
\end{figure}

The spectral fit, also for the \gr data, can be achieved with the values
$\eta=5\times10^{-4}$ of the proton injection rate and
$B_\mathrm{d}=142$~$\mu$G for the downstream magnetic field strength. Such
values of $B_\mathrm{d}$ are significantly higher than typical ambient magnetic
fields, also for the considered bubble wall in the dense environment (the gas
density at the shock is still a factor $\sim 100$ lower than in that
environment which implies a strongly decompressed field upstream of the SNR
shock compared to the value in the ambient medium). The high
$B_\mathrm{d}$-value must be attributed to field amplification at the shock
front due to the strong wave production by the acceleration of CRs far into the
nonlinear regime \citep{lucb00,bell04} \footnote{For direct observational
  evidence regarding magnetic field amplification in \rxj\, see
  \citet{vbk05,ballet06,bv06}.}

It is assumed that also electrons are injected into the acceleration process at
the shock front. Since the details of the electron injection process are poorly
known, the electron injection rate is chosen such that
the electron to proton ratio $K_\mathrm{ep}$ (which is defined as the ratio of
their distribution functions at all rigidities where the protons are already
relativistic and the electrons have not been yet cooled radiatively) is a
constant to be determined from the synchrotron observations. Clearly, from the
point of view of injection/acceleration theory, 
$K_\mathrm{ep}$, together with $B_0 \equiv B_\mathrm{d} /\sigma$ and $\eta$,
must be treated as theoretically not very well constrained parameters to
be {\it quantitatively} determined by comparison with the available synchrotron
and \gr data. Here $\sigma$ is the overall shock compression ratio.

In the present case the following parameter values are obtained by iteration:
$\eta=5\times10^{-4}$, $B_0=25$~$\mu$G, $K_\mathrm{ep}=1.3\times 10^{-3}$. The
gas dynamic variables at the present epoch are $\sigma=5.7$,
$\sigma_\mathrm{s}=3.3$, and $V_\mathrm{s}\approx 2200$~km sec$^{-1}$, where
$\sigma_\mathrm{s}$ and $V_\mathrm{s}$ denote the subshock compression ratio
and the overall shock velocity, respectively (Fig.1).

The corresponding solutions of the dynamic equations at each instant of time
yield the CR spectrum and the spatial distributions of CRs and thermal gas, and
therefore also the expected fluxes of nonthermal emission produced by the
accelerated CRs.

\begin{figure*}
\centering
\includegraphics[width=15.cm]{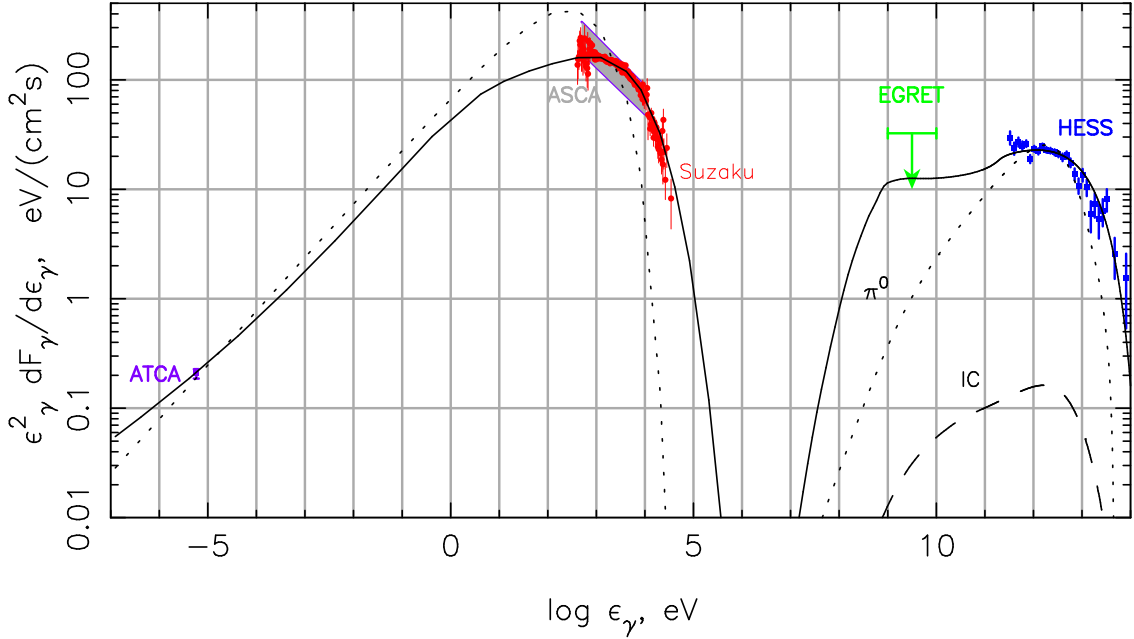}
\caption{Spatially integrated, overall nonthermal spectral energy distribution
  of \rxj\. The {\it solid} curve at energies above $10^7$~eV corresponds to
  $\pi^0$-decay \gr emission, whereas the {\it dashed} curve indicates the
  Inverse Compton (IC) emission.  The dotted line corresponds to the test
  particle limit which implies insignificant proton acceleration and magnetic
  field amplification \citep[see][for the details]{bv08}.  The ATCA radio data,
  as derived by \citet{acero09}, the ASCA X-ray data \citep[cf.][]{aha06}, the
  Suzaku X-ray data \citep{uat07}, and the 2006 HESS \gr data \citep{aha07} are
  also shown. The EGRET upper limit for the \rxj position \citep{aha06} is
  included as well. }
\label{f2}
\end{figure*}

The overall broadband spectral energy distribution (SED), expected at the
current evolutionary epoch, is displayed in Fig.2, together with the
experimental data from {\it ATCA} at radio wavelengths, as estimated for the
full remnant by \citet{acero09}, the X-ray data from {\it ASCA} \citep{aha06})
and Suzaku \citep{uat07}, and the H.E.S.S. \gr data \citep{aha07}.

The comparison with our earlier theoretical spectrum \citep{bv08} shows that
also a model with full spherical symmetry can successfully describe the
spatially integrated nonthermal emission properties of this object.

At energies $\epsilon_{\gamma}<1$~TeV the theoretical \gr spectrum is as hard
as $dF_{\gamma}/d\epsilon_{\gamma}\propto \epsilon_{\gamma}^{-1.8}$, whereas
for $\epsilon_{\gamma}>10$~TeV it has a smooth cutoff. It should be noted that
the \gr cutoff energy $\epsilon_{\gamma}^{\mathrm{max}}\approx
0.1cp_{\mathrm{max}}$ is sensitive to the magnetic field strength
$B_\mathrm{d}$, since the proton cutoff momentum has a dependence
$p_{\mathrm{max}}\propto R_\mathrm{s}V_\mathrm{s}B_\mathrm{d}$ \citep{ber96}.

The theoretical results presented here are fully consistent with a dominantly
hadronic origin of the observed TeV \grs. At the present epoch the SNR has
already converted $\approx 35\%$ of $E_\mathrm{sn}$ into accelerated nuclei.

\subsection{Estimate of the thermal X-ray emission}
Like for the case of RX~J0852.0-4622 \citep{bpv09} in addition the expected
flux of thermal X-rays is estimated here. This is only possible in a very
approximate way, even if we shall disregard the ejecta emission with the
argument that the ejected mass in the present phase is small compared to the
swept-up mass. The reason is the explosion into the wind bubble, creating a
rather different thermodynamical structure from that of a classical Sedov
solution for a SN explosion into a uniform medium, exclusively until now
considered in the literature. Yet the remnant of \rxj is well past the sweep-up
phase and has entered a quasi-Sedov phase in the stellar wind shell, i.e. a
roughly self-similar evolutionary phase, modified by strong particle
acceleration relative to a purely gas dynamic evolution. Other estimates of the
thermal emission from \rxj have been made by \citet{Katz08} and \citet{damg08},
effectively assuming an explosion into a uniform medium.

The approximation adopted here is the following.  The present bubble case is
compared with a SNR in a uniform medium in the classical Sedov phase without
any CR acceleration, making four assumptions (i) the total hydrodynamic
explosion energy is the same in both cases (ii) the shock velocity is the same
(iii) the present gas density upstream of the shock is the same, and (iv) the
two objects are at the same distance of 1 kpc. Then the results of
\citet{hsc83} for the thermal X-ray flux from a classical Sedov SNR are used,
employing the emission measure of the bubble remnant instead of that of the
classical Sedov remnant with the same four parameters above. This means that
the X-ray emissivity of the remnant is reduced by the ratio $R_\mathrm{em}
=EM_\mathrm{b}/EM_\mathrm{S}$ of the emission measure $EM_\mathrm{b}$ for our
bubble solution to the emission measure for the classical Sedov solution
$EM_\mathrm{S}$ which corresponds to a uniform ambient gas density
$N_\mathrm{g}(R_\mathrm{s})$ \citep{bpv09}.

The relation
\begin{equation}
  \frac{T_\mathrm{sub}}{T_\mathrm{s}}=
  \frac{[(\sigma_\mathrm{s} -1)(\gamma+1)+2](\gamma+1)^2}
{4 \gamma (\gamma -1) \sigma^2}
\end{equation}
is used, with $\gamma = 5/3$, to determine the subshock temperature
$T_\mathrm{sub}$, relevant for the thermal emission flux for the modified
shock.  $T_\mathrm{s}=10^7 (V_\mathrm{s}/(839 \mathrm{km/s}))^2 \mathrm{K}
  \approx 6.9 \times 10^7$~K denotes the gas temperature of an unmodified shock
  with $V_\mathrm{s} \approx 2200$~km/s.

Using the differential thermal X-ray model spectra $dF/d\epsilon\approx 3
\times 10^{-5}$~photons/(keV cm$^2$s) from \citet{hsc83} (for $T_\mathrm{e}
\neq T_\mathrm{i}$, see their Fig.2) for their $\eta = N_\mathrm{H}^2
E_\mathrm{sn} = 10^{49}$~erg cm$^{-6}$ with account of the scaling
$dF/d\epsilon\propto \eta $ with our model numbers $\eta \approx 8\times
10^{49}$~erg cm$^{-6}$, multiplying it by the factor $\theta^2
[E_\mathrm{sn}/(10^{51}\mathrm{erg})]^{-1/2}$, as required, where $\theta =
54.5$~arcmin for the angular size of the classical Sedov remnant corresponding
to \rxj\, and multiplying it also by the factor $R_\mathrm{em}=0.46$, results
in a thermal spectral energy density $\epsilon^2 dF/d\epsilon \approx 288$~eV
cm$^{-2}$ sec$^{-1}$ for $\epsilon =1$~keV.

This result has not yet taken into account that the actual postshock gas
temperature $T_\mathrm{sub}$ is by a factor $\approx 0.41$ lower than the
temperature $T_\mathrm{s}$ used in the above estimate. As discussed in
\citet{bpv09} this roughly corresponds to a reduction of the parameter value
$\eta \approx 3.3 \times 10^{49}$~erg cm$^{-6}$ and thus to a reduction of the
thermal flux at 1 keV to $\epsilon^2 dF/d\epsilon \approx 118$~eV cm$^{-2}$
sec$^{-1}$.

These numbers must be compared with the observed nonthermal X-ray energy flux
at 1 keV (see Fig.1) $\epsilon^2 dF/d\epsilon \approx 200$~eV cm$^{-2}$
sec$^{-1}$. Therefore, at 1~keV, the thermal energy flux comes out to be
smaller than the nonthermal flux by a factor of about 2. This does not
contradict the non-detection of thermal X-rays from \rxj.

\begin{figure}
\centering
\includegraphics[width=7.5 cm]{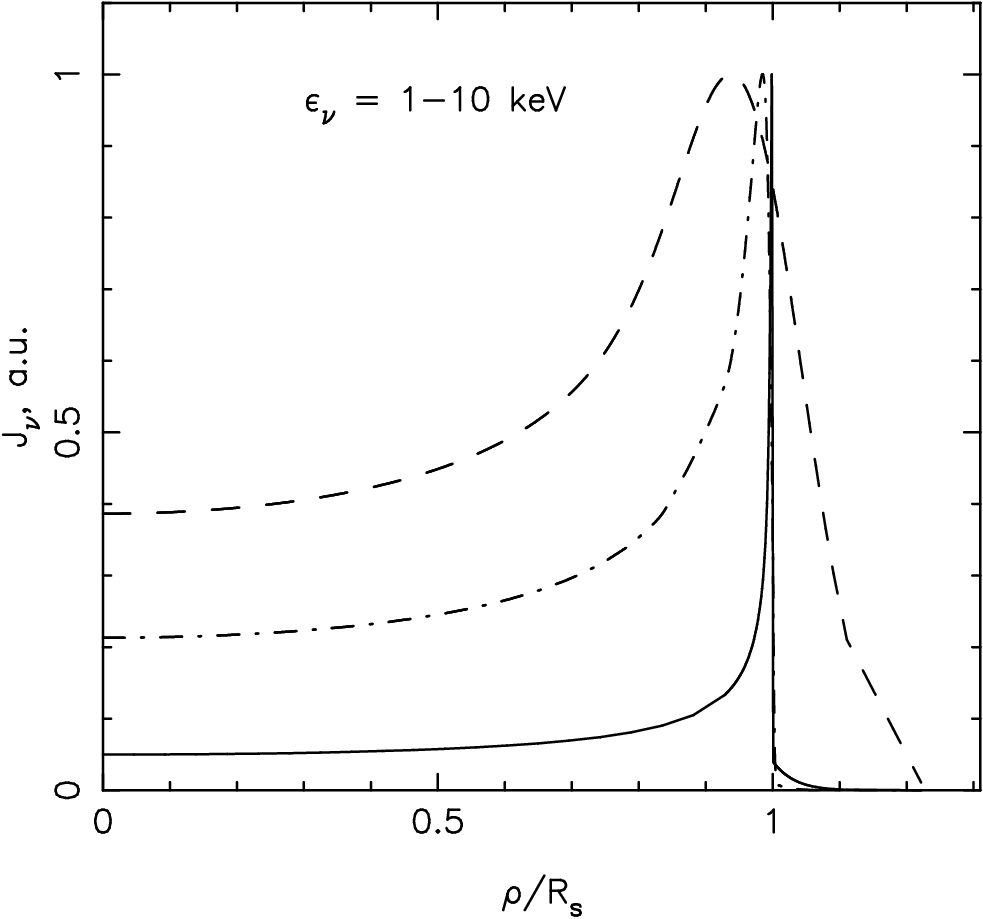}
\caption{ The X-ray emissivities for the energy
$\epsilon_{\gamma}=1-10$~keV as a function of projected, normalized
radial distance $\rho/R_\mathrm{s}$.The calculated radial profile is
represented by the {\it solid line}. The {\it dashed line}
represents the calculated profile convolved with the point spread
function of Gaussian width $0.05^\circ$. }
\label{f3}
\end{figure}
\begin{figure}
\centering
\includegraphics[width=7.5 cm]{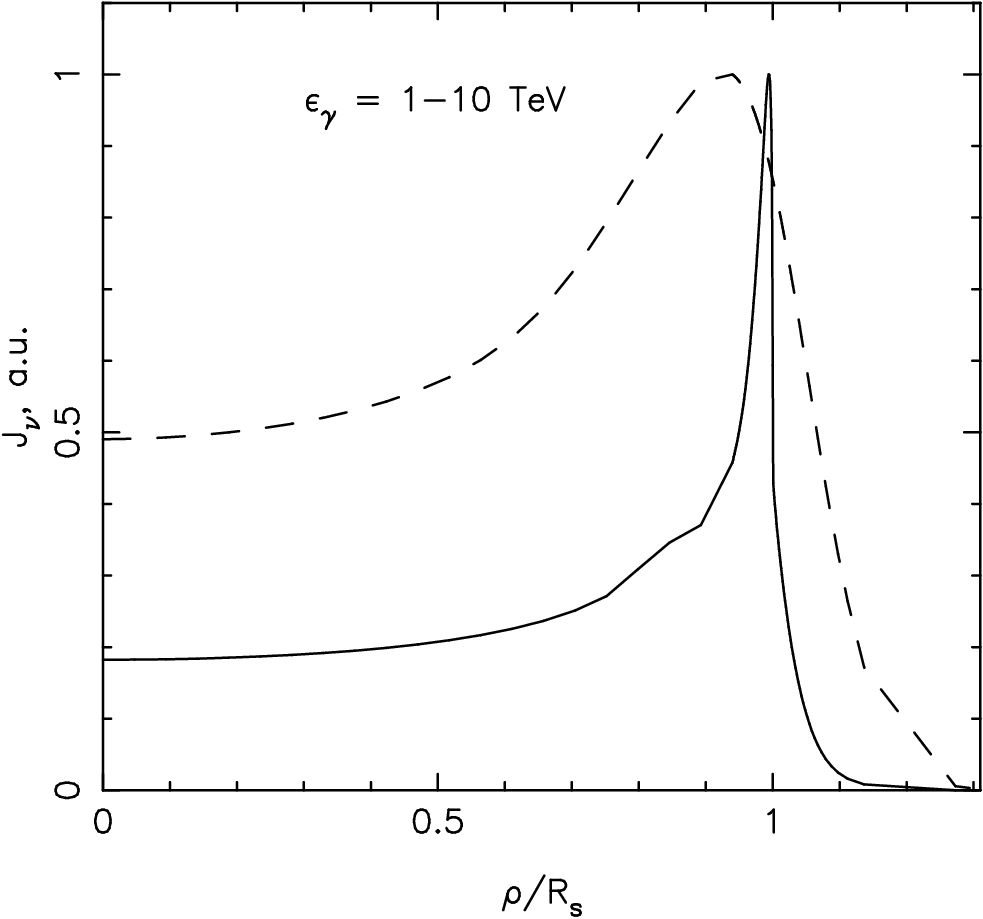}
\caption{ The \gr\ emissivities for the \gr\ energy
$\epsilon_{\gamma}=1-10$~TeV as a function of projected, normalized
radial distance $\rho/R_\mathrm{s}$. The calculated radial profile is
represented by the {\it solid line}. The {\it dashed line}
represents the calculated profile convolved with the point spread
function of Gaussian width $0.05^\circ$. }
\label{f4}
\end{figure}

\subsection{Spatial correlation of the nonthermal X-ray fluxes with the \gr
  fluxes}
Note that \citet{acero09}, who find a positive spatial nonlinear correlation
$F_\mathrm{X}\propto F_{\gamma}^{\alpha}$ with $\alpha=2.41$ between the flux
$F_{\gamma}$ of \grs with energies $\epsilon_{\gamma} = 1-10$~TeV and the flux of
nonthermal X-rays $F_\mathrm{X}$ with energies $\epsilon_{\gamma}=1-10$~keV, concluded
that such a correlation is in favor of a leptonic nature of the TeV-emission.
Such a positive correlation is naturally expected within the present approach.

\subsubsection{Correlation of the radial profiles}
In terms of the spherically symmetric model, discussed here, the 
correlation between X-ray and \gr fluxes means that the projected radial
profiles $J_\mathrm{X}(\rho)$ and $J_\mathrm{\gamma}(\rho)$ of these emissions
as functions of projected radial distance $\rho$ have a similar shape. At first
sight this is a contradiction to this model: as can be seen from Figs.3 and 4,
the calculated profiles are significantly different. Both profiles have a sharp
peak just behind the shock front, but the X-ray profile is by a factor of about
3 thinner than the \gr profile. However, when smoothed with the point spread
function of width $\sigma_0 =0.05^{\circ}$, corresponding to the
H.E.S.S. angular resolution, these profiles become very similar.
\citep{acero09} compared XMM measurements of the X-ray emission smoothed with a
point spread function of width $\sigma =0.05^{\circ}$. It can therefore be
concluded that the similarity of the X-ray and the \gr radial profiles is a
trivial consequence of the smoothing of two differently sharp peaks by a point
spread function whose width is considerably larger than the widths of either
one of the emission peaks.

The experimentally observed correlation according to Fig.9 of \citet{acero09}
is rather close to the relation $F_\mathrm{X}= F_{\gamma}$ for all the brighter
SNR regions.  Some deviation from this relation is observed for six of the
outer and dimmer regions, for which $F_\mathrm{X}< F_{\gamma}$. This
peculiarity can be also explained within the present model. The reason is that
protons with an energy of about 100~TeV occupy a region, which is noticeably
larger than the shock size $R_\mathrm{s}$, due to their high diffusive
mobility. The 10~TeV \gr emission which these protons produce has an even
larger radial extension since the gas density strongly increases with radial
distance. Such an effect is absent for the electrons of the same energy which
produce the nonthermal keV-emission: as a result of their strong energy losses
they occupy a region thinner than $10^{-2}R_\mathrm{s}$ around the shock
front. Therefore, as can be clearly seen from Figs.3 and 4, the ratio $
F_{\gamma}/F_\mathrm{X}$ is expected to increase as a function of radial
distance for $\rho/R_\mathrm{s}>1$, as the data of \citet{acero09} are
interpreted here.

\subsubsection{Correlation of azimuthal variations}

The measured X-ray and \gr emissions also undergo rather 
closely correlated azimuthal variations across the remnant. Such an effect
can not be described within the present model model, because it is spherically
symmetric. One can nevertheless attempt to interpret the 
correlated behaviour of the fluxes $F_\mathrm{X}$ and $F_{\gamma}$ in
different azimuthal sectors of the remnant. It is natural to assume that this
variation of the remnant properties is due to variations of the ambient gas
density $\varrho$.  According to the model proposed here the SNR's present
evolutionary phase is intermediate between the sweep-up and a quasi-Sedov
phase. Since the ejecta kinetic energy together with the kinetic energy of the
swept-up gas contain about half of the explosion energy, the SN shock can be
treated approximately as a piston-driven shock. The speed of such a shock
depends only weakly upon the upstream gas density. Therefore $V_\mathrm{s}$ can
be approximated by a constant across the whole remnant.

The local hadronic \gr emission then varies like $F_{\gamma}\propto
\varrho E_\mathrm{c}\propto \varrho$, where the local CR energy content
$E_\mathrm{c}$ in each angular sector is assumed to be the same everywhere,
because the overall CR energy content $E_\mathrm{c}\approx 0.4E_\mathrm{SN}$
has already reached the saturation level and therefore can not undergo
significant variations.  In addition, the highest energy CRs in the SNR
have the highest diffusive mobility so that they tend to be distributed roughly
uniformly across the remnant. 

As it is clear from Fig.2, the synchrotron spectrum at high energies
$\epsilon_{\gamma}>10$~eV is already considerably influenced determined by
synchrotron losses of the emitting electrons. Since the energy of these
electrons is rather rapidly and completely transformed into synchrotron
emission, the flux $F_\mathrm{X}\propto K_\mathrm{ep}\varrho V_\mathrm{s}^3$ is
determined by the available energy flux $\rho V_s^3/2$ into the shock, an
essential part of which is transformed at the shock front into the energy flux
of high energy CRs. High energy electrons therefore accumulate a fraction
$K_\mathrm{ep}$ of this flux. This flux hardly
depends on the value of the interior magnetic field in all those SNR regions
where the field is sufficiently high, $B_\mathrm{d}\gsim 140$~$\mu$G.  In this
case the relation $F_\mathrm{X}\propto F_{\gamma}$ is expected for the
brightest part of remnant.  This is indeed observed \citep{acero09}.  Since the
energy density of the amplified magnetic field $B^2_\mathrm{d}$ is
phenomenologically known to be proportional to 
$\varrho V_\mathrm{s}^{\beta}$, with $ 2 \leq \beta \leq 3$ \citep{vbk05},
it is expected to be proportional to the local upstream gas density
$\varrho$. The magnetic field strength is therefore lower than the average
value $\approx 140$~$\mu$G within those parts of the remnant, where the gas
density is lower than the average value. Synchrotron losses become much smaller
in these regions and therefore the expected synchrotron X-ray flux is
approximately $F_\mathrm{X}\propto \varrho
K_\mathrm{ep}B_\mathrm{d}^{3/2}\propto \varrho^{7/4}$.  For the dim part of the
remnant, in which X-ray and \gr emission is considerably lower than on average,
one then obtains $F_\mathrm{X}\propto F_{\gamma}^{7/4}$. This is approximately
consistent with the observational result.

\section{Summary}
The assumption that CR injection/acceleration takes place uniformly
across the whole SN shock surface is consistent with the existing data.

The swept-up mass is so small in this case that, in a rough approximation, the
estimated flux of thermal X-rays at 1 keV is lower than the nonthermal flux,
consistent with the nondetection of thermal X-ray emission until now.

It is concluded that the present observational knowledge of SNR \rxj can be
interpreted by a source which ultimately converts more than 35\% of the
mechanical explosion energy into nuclear CRs.  Also the observed high energy
\gr emission of SNR \rxj turns out to be primarily of hadronic origin.

\begin{acknowledgements}
The authors thank V.S. Ptuskin and V.N. Zirakashvili for discussions on
    the thermal emission properties. This work has been supported in part by
  the Russian Foundation for Basic Research (grants 06-02-96008,
  07-02-0221). EGB acknowledges the hospitality of the Max-Planck-Institut
  f\"ur Kernphysik, where part of this work was carried out.
\end{acknowledgements}


\begin{thebibliography}{99}

\bibitem[Abbot(1982)]{abb}
Abbott, D.C. 1982, ApJ, 263, 723

\bibitem[Acero et al.(2009)]{acero09}
  Acero, F., Ballet, J., Decourchelle, A., et al. 2009, to be published in
  A\&A; arXiv:0906.1073v1 [astro-ph.HE]

\bibitem[Aharonian et al.(2004)]{aha04}
Aharonian, F. A., Akhperjanian, A., Aye, K.-M., et al.(H.E.S.S.
Collaboration) 2004, Nature, 432, 75

\bibitem[Aharonian et al.(2006)]{aha06}
Aharonian, F. A., Akhperjanian, A., Bazer-Bachi, A.R., et al. (H.E.S.S.
Collaboration) 2006, A\&A, 449, 223

\bibitem[Aharonian et al.(2007)]{aha07}
Aharonian, F. A., Akhperjanian, A., Bazer-Bachi, A.R., et al. (H.E.S.S.
Collaboration) 2007a, A\&A, 464, 235

\bibitem[Ballet(2006)]{ballet06}
Ballet, J. 2006, Adv. Space Res., 37, 1902

\bibitem[Bell(2004)]{bell04}
Bell, A. R. 2004, MNRAS, 353, 550

\bibitem[Berezhko(1996)]{ber96}
Berezhko, E.G. 1996, Astropart.\ Phys., 5, 367

\bibitem[Berezhko et al.(1996)]{byk96}
Berezhko, E.G., Elshin, V.K. \& Ksenofontov, L.T. 1996 JETPh, 82, 1

\bibitem[Berezhko \& V\"olk(2000)]{bv00}
Berezhko, E.G. \& V\"olk, H.J. 2000, A\&A, 357, 183

\bibitem[Berezhko(2005)]{ber05}
Berezhko, E. G. 2005, Adv. Space Res., 35, 1031

\bibitem[Berezhko(2008)]{ber08}
Berezhko, E.G. 2008, Adv. Space Res., 41, 429

\bibitem[Berezhko \& V\"olk(2006)]{bv06}
Berezhko, E.G. \& V\"olk, H.J. 2006, A\&A, 451,981

\bibitem[Berezhko \& V\"olk(2008)]{bv08}
Berezhko, E.G. \& V\"olk, H.J. 2008, A\&A, 492,695

\bibitem[Berezhko et al.(2009)]{bkv09}
  Berezhko, E.G., Ksenofontov, L.T., \& V\"olk, H.J. 2009, A\&A, in press;
  arXiv:0906.3944v1 [astro-ph.HE]

\bibitem[Berezhko et al.(2009)]{bpv09}
Berezhko, E. G., P\"uhlhofer, G., \& V\"olk, H. J. 2009, A\&A in press;
arXiv:0906.5158v1 [astro-ph.HE]

\bibitem[Cassam-Chena\"i et al.(2004)]{cassam04} 
Cassam-Chena\"i, G., Decourshelle, A., Ballet, J., et al. 2004, A\&A, 427, 199

\bibitem[Drury et al.(2008)]{damg08} Drury, L.O'C., Aharonian, F.A., Malyshev,
  D. \& Gabici, S. 2008, A\&A, 496, 1

\bibitem[Enomoto et al.(2002)]{enomoto02}
Enomoto, R., Tanimori, T., Naito, T., et al. 2002, Nature, 416, 823

\bibitem[Enomoto et al.(2006)]{enomoto06}
Enomoto, R., Watanabe, S., Tanimori, T. et al. 2006, ApJ, 652, 1268

\bibitem[Fukui et al.(2003)]{Fukui}
Fukui, Y., Moriguchi , Y., Tamura, K. et al. 2003, PASJ, 55, L61

\bibitem[Hamilton et al.(1983)]{hsc83}
Hamilton, A.J.S., Sarazin, C.L. \& Chevalier, R.A. 1983, ApJS, 41, 115

\bibitem[Hiraga et al.(2005)]{Hiraga}
Hiraga, J.S., Uchiyama, Y., Takahashi, T., et al. 2005, A\&A, 431, 953

\bibitem[Katz \& Waxman(2008)]{Katz08} 
Katz, B. \& Waxman, E. 2008, JCAP, 01, 1

\bibitem[Koyama et al.(1997)]{koyama97}
Koyama, K., Kinagasa K., Matsuzaki, K., et al. 1997, PASJ, 49, L7

\bibitem[Lazendic et al.(2004)]{laz04}
Lazendic, J. S., Slane, P.O., Gaensler, B.M., et al. 2004, ApJ, 602, 271

\bibitem[Lucek \& Bell(2000)]{lucb00}
Lucek, S.G. \& Bell, A.R. 2000, MNRAS, 314, 65

\bibitem[Moriguchi et al.(2005)]{mor05}
Moriguchi, Y., Tamura, T., Tawara, Y., et al. 2005, ApJ, 641, 947

\bibitem[Muraishi et al.(2000)]{mur00}
Muraishi, H., Tanimori, T., \& Yanagita, S. 2000, A\&A, 354, L57

\bibitem[Pfeffermann \& Aschenbach(1996)]{pfe96}
Pfeffermann, E. \& Aschenbach, B. 1996, in: R\"ontgenstrahlung from the
Universe. (Eds. H.U. Zimmermann, J. Tr\"umper \& H. Yorke (MPE Rep. 263,
Garching), 267

\bibitem[Plaga(2008)]{Plaga}
Plaga, R. 2008, New Astronomy, 13, 73

\bibitem[Porter et al.(2006)]{Porter06}
Porter, T.A., Moskalenko, I.V., Strong, A.W. 2006, ApJ, 648, L29

\bibitem[Slane et al.(1999)]{sla99}
Slane, P., Gaensler, B. M., Dame, T. et al. 1999, ApJ, 357, SL99

\bibitem[Takahashi et al.(2008)]{ttu08}
Takahashi, T., Tanaka, T. Uchiyama, Y., et al. 2008, PASJ, 60, S131

\bibitem[Tanaka et al.(2008)]{tua08}
Tanaka, T., Uchiyama, Y., Aharonian, F.A., et al. 2008, to appear in ApJ,
[arXiv0806.1490 (astro-ph)]

\bibitem[Uchiyama et al.(2007)]{uat07}
Uchiyama, Y., Aharonian, F.A., Tanaka, T., et al. 2007, Nature, 449, 576

\bibitem[V\"olk et al.(2003)]{vbk03}
V\"olk, H.J., Berezhko, E.G. \& Ksenofontov, L.T. 2003, A\&A, 409,
563

\bibitem[V\"olk et al.(2005)]{vbk05}
V\"olk, H.J., Berezhko, E.G. \& Ksenofontov, L.T. 2005, A\&A, 433,
229

\bibitem[Wang et al.(1997)]{wang97}
Wang, Z.R., Qu, Q.-Y., \& Chen, Y. 1997, A\&A, 318, L59

\bibitem[Zirakashvili(2009)]{zirak09}
Zirakashvili, V.N. 2009, in ``High Energy Gamma-Ray Astronomy''
(Eds. F.A. Aharonian, W. Hofmann, F.M. Rieger), Melville, New York, 2009, AIP
Conf. Proc. 1085, p. 129 ff.


\end{thebibliography}
\end{document}